# Recent advances in optics (RAO): Contribution of D Narayana Rao in optics and photonics research in India

Jitendra Nath Acharyya and G Vijaya Prakash

*Nanophotonics Lab, Department of Physics, Indian Institute of Technology Delhi,*

*New Delhi- 110 016, India,*

Dedicated to Professor D N Rao for his significant contributions and pioneering works in the fields of spectroscopy, optics, nonlinear optics and photonics.

The first documented speculations on light were found in oriental and Greek schools of philosophy. The investigations on the light in India were rooted way back to the ancient era. The Indian Samkhya, Nyaya, and Vaisheshika identified light or fire (*Tejas*) as one of the key elements among the five elementary things of the universe. The pioneering results of optics and spectroscopy were witnessed in the era of stellar people like Sir Jagadish Chandra Bose and Sir C V Raman. The research has then taken a new directive into the development of nonlinear optics and photonics after the invention of the laser. The present article reviews the past few decades pioneering works of Prof. D Narayana Rao, an experimental physicist, in the context of optics and photonics in India. The most notable contributions of Prof Rao, introduced in India for the first time, are *white-light interferometry, degenerate four-wave mixing* (DFWM), *electric-field induced second harmonic generation* (EFISHG), *incoherent laser spectroscopy* (using dye laser), and *femtosecond lasers for creating nano/microstructures*. © Anita Publications. All rights reserved.

## 1 Introduction

The notion of light rooted in the ancient culture was religious. The bible describes the light as divine things with the famous quotation *"Let there be light,"* and the light was separated from the darkness at the genesis of the universe. The field of optics deals with the properties and different phenomenon of light. *'Photonics'* has emerged after the invention of the laser and developed rapidly for the upliftment of modern human society. The twenty-first century is called the era of photonics, which underpins technologies starting from the smartphones, internet, telecommunications to medical purposes. Today, India has evolved as one of the leaders in optics and photonics in the world. One of the leading optics and photonics researchers of India is Prof D Narayana Rao (hereafter referred to as DNR) of the University of Hyderabad, who took experimental optics in an entirely new direction. The current article is dedicated to DNR, and covers briefly his broad spectrum of research areas, mostly contributed by him, his students, and collaborators. In 1990, Prof DNR joined as a Reader at the University of Hyderabad and introduced for the first time in India several new experimental techniques related to optics, exclusively their-order related nonlinear optical techniques, namely *white-light interferometry, degenerate four-wave mixing* (DFWM), *Z-scan*, *electric-field induced second harmonic generation* (EFISHG), *incoherent laser spectroscopy* (using dye laser), and *femtosecond lasers for creating nano/microstructures*. Prof Rao's extensive contributions gave a new direction to optics

---

*Corresponding author*
*e-mail: jitendraphy12@gmail.com* (Jitendra Nath Acharyya); *prakash@physics.iitd.ac.in* (G Vijaya Prakash)

research in Indian laboratories, especially to encourage non-premier organisations like state universities. The other pioneering contributions are femtosecond supercontinuum generation, local field effects in dielectric media, optical limiting and switching in nanomaterials, electroabsorption spectroscopy, and photonic crystals and waveguides for various applications. It was extremely difficult to include all the research works done by DNR in a short review, therefore, we briefly outlined here topics of his interest and innovative ideas cultivated over the past three decades of his tireless research activities.

The following sections are categorized depending on the research field with a brief introduction about the topic. In some cases, the proper chronological orders were not maintained to draw a pictorial view of all the reach areas covered in this review article.

## 2 Pancharatnam phase and spectral interferometry

The University of Hyderabad is one of the pedestals of optics and photonics research in India. DNR established an optics and photonics laboratory at the University of Hyderabad equipped with a variety of experimental setups related to experimental optics. White-light spectral interferometry research, for the first-time in India, was started by DNR, in 1990. The initiation of such an idea was born out of the simplest technique based on the Pancharatnam phase [1]. The wavefunction of a quantum mechanical system undergoes a phase shift due to the cyclic variation of the system paraments before coming to their original values. This phase is known as the geometrical phase or Berry phase [1]. The optical analog of the Berry phase is known as the Pancharatnam phase, which can be generated by the cyclic variation of the state of polarization of a light beam. The demonstration of such a Pancharatnam phase can be achieved by simple white-light interferometry [based on Mach – Zehnder (MZ) configuration] as shown in the schematic diagram Fig 1(a). An automobile lamp bulb was used as a source even it can be performed with low power, He-Ne laser. The beam splitter (BS) divides the beam into two paths, which produce the interference fringes at the back focal plane of lens L3. Using this simple setup, the interference (both in MZ and Michelson configurations) fringes can be easily obtained with a white light source. The interference fringes are extremely stable and free from the air current disturbance. Both MZ and Michelson interferometry [2-7] designs are shown in Figs 1(a) and 1 (b).

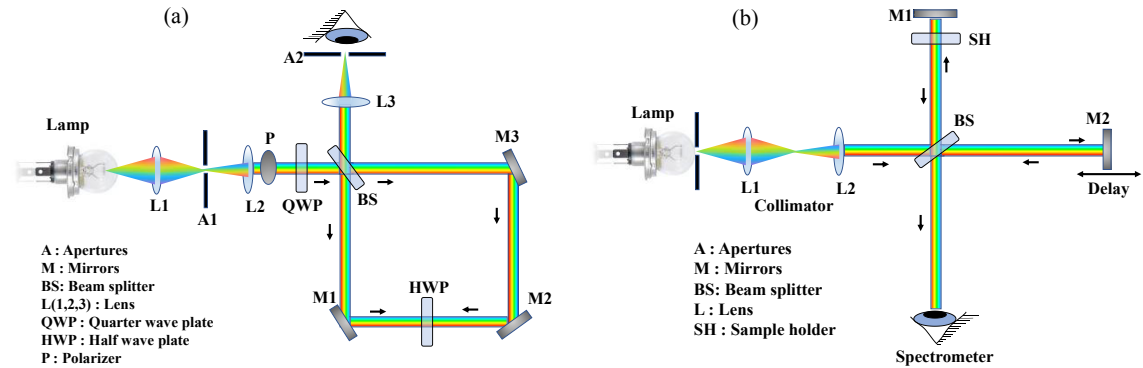

Fig 1. (a) Schematic arrangements of the white-light interferometry for demonstration of the Pancharatnam phase [1]. (b) Schematic experimental setup for refractive index measurements using Michelson white-light interferometry [5].

The spectral interferometry is based on the law of spectral interference, which can be expressed as follows

$$S(\lambda) = \frac{1}{2} S_0(\lambda) \left[ 1 + \text{Re}[\mu_{12}(\lambda)] \cos\left(\frac{2\pi\Delta}{\lambda} + \theta\right) \right] \quad (1)$$

where, $\mu_{12}(\lambda)$ represents the spectral coherence of the two beams at wavelength $\lambda$ with a path difference of $\Delta = n(\lambda)2t - L_0$, where $t$ is the thickness and $L_0$ accounts for the constant path difference between the two beam arms to obtain the zero-order at the center. $n(\lambda)$ is the wavelength-dependent refractive index.

In the early 90's, Nirmal Kumar *et al* reported a drastic change in the interference pattern when a dispersive medium is inserted into one of the interferometer arms of the Michelson white-light interferometer (Fig 1b) [2-4]. Later, Prakash *et al* demonstrated linear optical properties of different glass materials using such novel white-light interferometry technique of miniature form utilising fiber optic spectrometers and short lens/mirror posts. Figure 1(a) shows the schematic of the Michelson type white-light interferometer for refractive index measurements. The white-light interferometry can explain many intriguing optical phenomena such as spectral anomalous and refractive index dispersion and so on [5-7].

## 3 Incoherent laser spectroscopy using home-built broadband dye laser

Nowadays, the pulsed (nanoseocnd to femtosecond) lasers are abundant in most of the Indian labs whereas, it was scarcely available in the early 90s. In 1990, for the first time in India, DNR introduced the idea of incoherent laser spectroscopy using a home-built dye laser as an incoherent broadband laser source. The spectral modulations on a broadband spectral range were studied using a home-built broadband dye laser source [3,8]. The gain medium was Rhodamine B (RhB) in methanol, which was continuously circulated to minimize the laser scattering and decompositions. The spectral bandwidth was about $\Delta\lambda$ ~80nm, which is equal to 168 fs. The home-built dye laser system consists of an oscillator and an amplifier. In a single-stage amplification process, the output of the oscillator was amplified. A part of the power from the second harmonic of an Nd:YAG laser (532 nm, 10 Hz, 6 ns FWHM, 100 mJ/pulse) was used to pump the oscillator, and the other remaining was used for pumping the amplifier. Figure 2(a) shows the schematic of the home-built broadband dye laser system [8,9]. This low-cost dye laser is a perfect replacement for the expensive femtosecond laser, and it has been well-utilised in the incoherent laser spectroscopy, which will be discussed later.

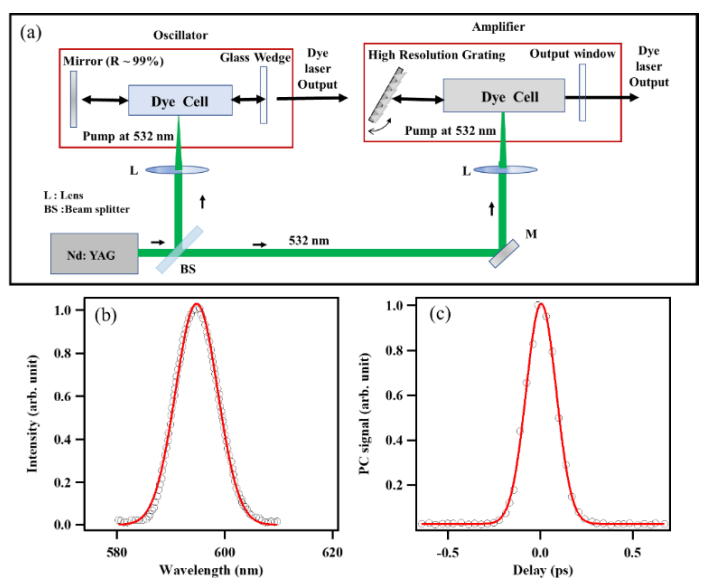

Fig 2. (a) Schematic of the home-built broadband dye laser system, (b) Emission Spectrum of RhB (in methanol). The open circles represent the experimental data, and the solid line is the Gaussian fit with a width of 7.8 nm, (c) Autocorrelation trace of time-resolved signal (circles) of RhB (in methanol). The solid line is the Gaussian fit with a width of about 168 fs [3,8].

In other contributions in the development of lasers, the notable work is the optically-pumped multigas Far-IR laser with different gases ($CH_3OH$, $CH_3Br$, $CH_3I$) in the Far-IR cavity. An efficient FIR lasing action was achieved on several lines of various gases by adjusting the partial pressure of the gases and the $CO_2$ laser line [10].

Being a nonlinear optics expert, DNR has imprinted his novel ideas in second-order and third-order optical studies. The following sections are based on the second-order and third-order nonlinear optical properties studied in both organic and inorganic molecular systems.

## 4 Electric field-induced second harmonic generation (EFISHG)

An exciting field initiated by DNR first time in India, is *electric field-induced second harmonic generation* (EFISHG) in the early 90s. The second-order effect is absent in centrosymmetric ($\chi^{(2)} = 0$) molecules such as gases, liquids, and solutions. If the molecules possess a permanent dipole moment, a partial alignment of the molecules is obtained. An external DC electric field application leads to symmetry breaking of centrosymmetric molecules resulting in efficient second harmonics generations by electric field induced second harmonic generation (EFISHG) technique. DNR designed a novel optical cell for the EFISHG measurements making use of a conventional spectrometer glass cuvette for different organic molecules in solution phases. Figure 3 represents the schematic experimental arrangements of EFISHG and the novel sample cell design.

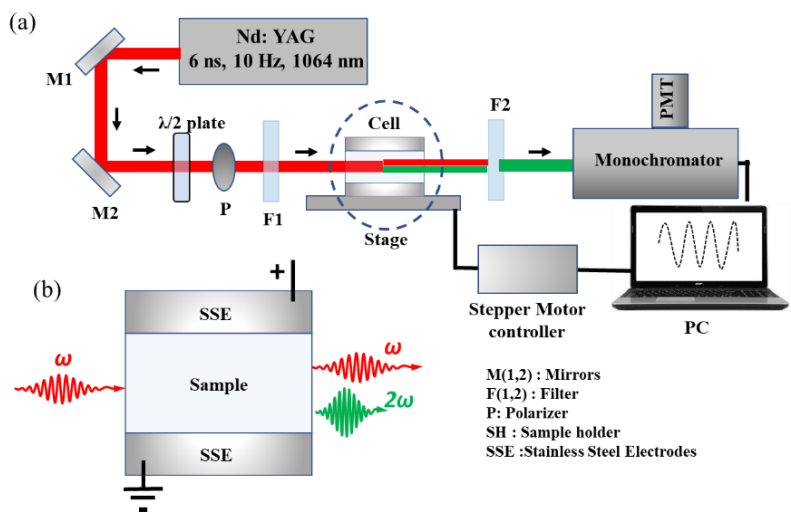

Fig 3. (a) Schematic experimental arrangements for EFISHG. (b) The simplified view of the novel cell designed for EFISHG experiment.

The intensity of the generated second harmonic in EFISHG technique can be expressed as follows,

$$I_{2\omega}^{EFISHG} = k_{inst} \left| \left(-\frac{\chi_1}{\Delta\varepsilon_1}\right) F(e^{-\alpha_{2\omega}L/2} - e^{i\phi(L)}) \right|^2 I_\omega^2 \, E_0 \tag{2}$$

where, $k_{inst}$ is the instrument constant. $F$ is the product of different Fresnel factors, and $\alpha_{2\omega}$ denotes the absorption coefficient of the liquid at the second harmonic frequency. The input laser intensity is $I_\omega$ with the applied DC electric field $E_0$. The dielectric constant dispersion is defined as $\Delta\varepsilon_1 = (n_{2\omega}^2 - n_\omega^2)$, with a periodic phase factor defined as, $\phi(L) = (\lambda/2L)(n_{2\omega} - n_\omega)$, where $L$ is the path length inside the cell. The first hyperpolarizability ($\beta$) can be determined using Eq (2) by fitting the experimental data as described in the literature [11]. The molecular hyperpolarizability can be experimentally determined by the EFISHG technique [11]. An efficient and traditional protocol was developed to analyze the EFISHG data using the internal cancellation of second-harmonic generation in solvent mixture and solutions. In 2004, the first hyperpolarizability of a zwitterionic push-pull molecule (DCNQI) was reported by DNR along with Prof T P Radhakrishnan of the School of chemistry [11]. Gangopadhyay *et al* [12] reported substituted *n*-alkyl groups as the potential non-centrosymmetric materials for efficient optical second-harmonic generation. Ravi *et al* [13,14] reported the influence of H-bonding on the solid-state second-harmonic generation of Chiral Quinoid compounds. The powder SHG is determined by the dual influence of H-bonding on the crystal structure and hyperpolarizability. Patil *et al* reported the second harmonic generation in chalcone derivatives [15].

### 5 Third-order nonlinear optical studies

DNR is also known in India for his pioneering contribution in third-order nonlinear optical studies, based on single beam nonlinearity (known as Z-Scan) and four-wave mixing. To our knowledge, for the first time In India, DNR established the experimental setup for the degenerate four-wave mixing (DFWM) technique using the incoherent laser source at the University of Hyderabad in the early 90s. In the following sections, a brief development of DFWM based *incoherent laser spectroscopy* was demonstrated.

The central description of all four-wave mixing processes is described by the interaction of three electromagnetic fields along with the generation of the fourth field. The first input electromagnetic field creates oscillating polarization in the medium. The application of the second field produces interaction with the first field, which generates the harmonics of the polarization. Now, the application of the third field causes the beat as well as the sum and difference frequency generation with both the other two input fields. This results in the generation of the fourth electromagnetic field in the four-wave mixing technique [16]. If the frequencies of the input electromagnetic fields are equal, it is called degenerate four-wave mixing (DFWM) geometry. A schematic experimental arrangement is given in Fig 4 (a). The fourth $k_4$ field generation depends on the phase-matching condition as $k_4 = k_3 - k_2 + k_1$. The third-order nonlinear susceptibility can be obtained with respect to a reference sample according to the following equation [17],

$$\chi^{(3)}_{sample} = \left(\frac{n_{sample}}{n_{ref}}\right)^2 \left(\frac{I_{sample}}{I_{ref}}\right)^{1/2} \left(\frac{L_{ref}}{L_{sample}}\right) \alpha L_{sample} \left(\frac{e^{(\alpha L_{sample})/2}}{1 - e^{-\alpha L_{sample}}}\right)^2 \chi^{(3)}_{ref} \tag{3}$$

where, '*I*' represents the DFWM signal intensity, $\alpha$ is called the linear absorption coefficient, *L* is the sample length, and '*n*' is called the linear refractive index.

In the early 80s, in the State University of New York at Buffalo, DNR was extensively involved in DFWM measurements in picosecond and subpicosecond regimes. It was demonstrated that the optical light modulations in bacteriorhodopsin films using a degenerate four-wave mixing geometry, acts as all-optical logic AND, and OR gates [18-20]. The degenerate four-wave mixing technique can be utilized to measure the third-order nonlinear optical susceptibility. This was first demonstrated by DNR *et al* from experimental observation of π-electron conjugated systems in picosecond time scaled nonlinear optical response [21].

In the early 90s, at the University of Hyderabad, the DFWM measurement was started using the home-built dye laser source [9], which was used as an incoherent laser source mimicking femtosecond laser pulses. Later on, the DFWM was performed from nano to femtosecond time domains. The population relaxation time of some organic dyes such as erythrosin B, commercial ink, and metalloporphyrin was first investigated using a DFWM technique with such dye-laser-based incoherent laser source [8,9]. Raavi *et al* reported the ultrafast nonlinear optical properties and ultrafast time response of different phthalocyanine samples using femtosecond DFWM measurements [17].

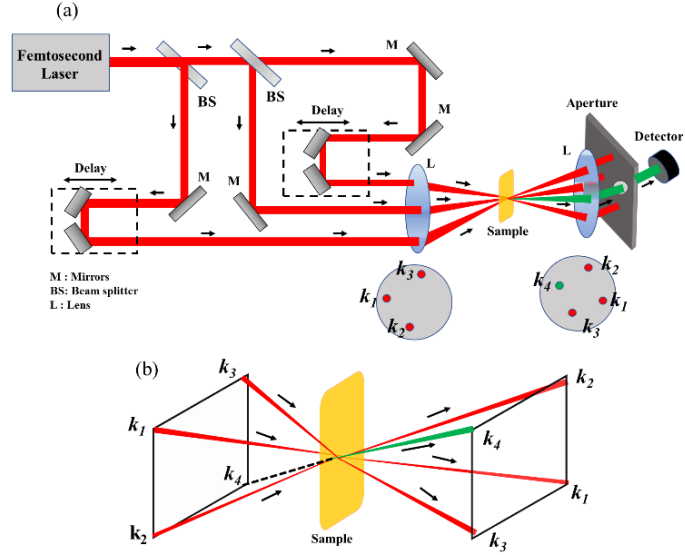

Fig 4. (a) Schematic of degenerate four-wave mixing (DFWM) experimental arrangements in BOXCARS geometry. (b) the schematic of BOXCARS geometry for the interactions of the three electromagnetic fields and generations of the fourth field in DFWM [17].

Another notable contribution of DNR is in single-beam laser nonlinearity, known as Z-scan. While Z-scan was first reported in 1994-96 in laser-based institutes (TIFR and RRCAT), DNR popularised and introduced Z-scan at the university level at a similar time. The following section demonstrates the underlying physics of Z-scan and some notable contributions of DNR. Third-order nonlinear optical properties of different nonlinear materials are characterized using Z-scan technique. In Z-scan measurements, a Gaussian laser beam is focused by a convex lens, and the transmittance through the sample is measured as the sample moves along the propagation direction of the beam. As depicted in Fig 4, all the transmitted beam through the sample is collected in an open aperture configuration. In a closed aperture configuration, an aperture is used before collecting the transmitted beam. The open aperture Z-scan trace conveys the nonlinear absorption information, whereas the nonlinear refraction can be extracted from the closed aperture trace. The Z-scan profiles are analyzed using Gaussian laser beam approximation,[22] with the electric field distributions for a $TEM_{00}$ as follows

$$E(z, r) = E_0 \left(\frac{\omega_0}{\omega(z)}\right) e^{-r^2/\omega^2(z)} \, e^{-ikr^2/2R(z)} \, e^{-i\varphi(z)} \quad (4)$$

where $E_0$ is the amplitude of the electric field at the focus. The beam waist ($w(z)$) at $z$ is related to the beam waist ($w_0$) at the focus as $w^2(z) = w_0^2 (1 + z^2/z_r^2)$, where $z_r$ is the Rayleigh length, $\lambda$ is the wavelength of the laser beam, $k = 2\pi/\lambda$ is the wave vector. $R(z) = z(1 + z^2/z_r^2)$ represents the radius of curvature of the wave-front. The term $\exp(-i\varphi(z))$ in the electric field, provides the radially uniform phase variations. In the presence of nonlinear absorption (under thin sample approximation, sample length $L \gg z_r$), the intensity and phase change within the sample can be expressed using slowly varying envelope approximations (SVEA) by a differential equation as [22]

$$\frac{dI}{dz'} = -\alpha(I)I \qquad \text{and} \qquad \frac{d\Delta\varphi}{dz'} = \Delta n(I)k \quad (5)$$

where $z'$ denotes propagation distance, $\alpha(I)$ includes all linear and nonlinear absorption terms, which can be expressed as $\alpha(I) = (\alpha_0 + \beta I + \gamma I^2 + \delta I^3 ...)$, where $\alpha_0$ (cm$^{-1}$) is the linear absorption coefficient. The nonlinear terms $\beta$ (cm/GW), $\gamma$ (cm$^3$/GW$^2$) and $\delta$ (cm$^5$/GW$^3$) are for two-photon absorption (2PA), three-photon absorption (3PA) and four-photon absorption (4PA) coefficients, respectively. The factor $\Delta n(I)$ represents the intensity-dependent refractive index change. The nonlinear phase-shift ($\Delta\varphi$) can be calculated as,

$$\Delta\varphi(z,r) = \left(\frac{\Delta\varphi_0}{1 + z^2/z_r^2}\right) \exp\left(\frac{2r^2}{w^2(z)}\right) \tag{6}$$

$\Delta\varphi_0$ is the on-axis phase shift at the focus which is related to the nonlinear refractive index ($n_2$) as $\Delta\varphi_0 = kn_2I_0 L_{eff}$, where the effective sample length, $L_{eff} = \dfrac{1 - e^{-\alpha_0 L}}{\alpha_0}$ and $L$ is the sample length. Finally, the Z-scan transmittance for far-field aperture can be solved using the 'Gaussian decomposition' method and finally obtained the transmittance equations as follows [22],

$$T(z) = \frac{\int_{-\infty}^{\infty} P_T(\Delta\varphi_0(t))}{S \int_{-\infty}^{\infty} P_i(t)\, dt} \tag{7}$$

$S = 1 - \exp(-2r_a^2/w_a^2)$ is called the aperture linear transmittance, where $r_a$ and $w_a$ denote the aperture radius and beam radius at the aperture, respectively. $P_T$ is the transmitted power through the aperture, and $P_i$ is the instantaneous input power. For open-aperture Z-scan, $S = 1$.

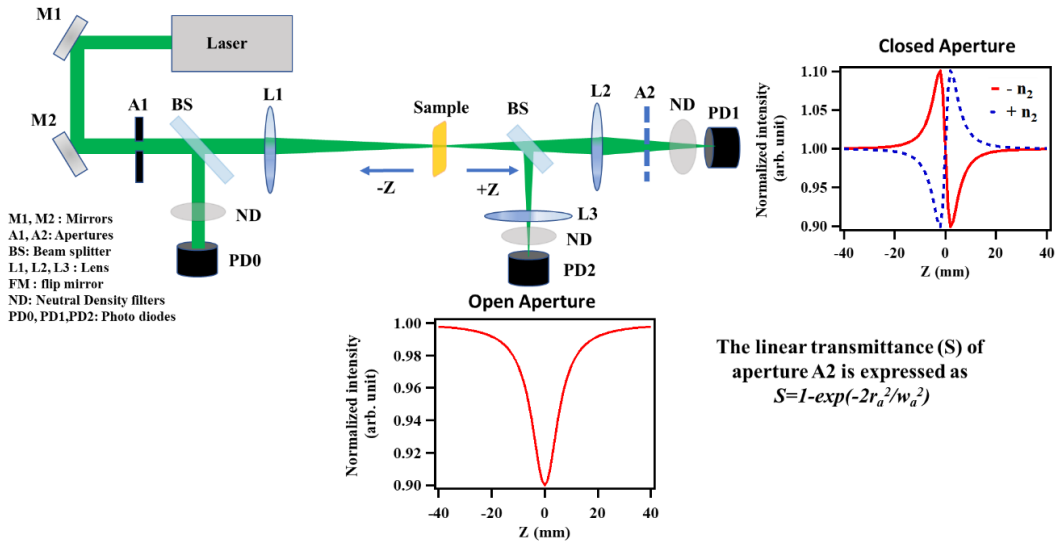

Fig 5. Schematic experimental arrangements of single-beam Z-scan measurement [23].

For almost two decades DNR *et al* investigated nonlinear optical properties of wide varieties of materials, noble metals to inorganic/organic materials, utilisiing diverse pulsed (nanosecond, picosecond, and femtosecond) lasers. In 1998, Rao *et al* reported the first comparative study on three important optical limiters, namely $C_{60}$, Phthalocyanine, and Porphyrin, with the dispersion behavior in the visible spectral regime [24]. Results clearly showed that Porphyrin is most suitable for optical limiting applications in the region 500-600 nm, whereas Phthalocyanine is suitable in the region 475-680 nm, and the $C_{60}$ is ideal in the region 430-650 nm. In 1998, Rao *et al* again reported the dispersion of nonlinear absorption behavior of $C_{60}$ using open aperture Z-scan over the visible spectral region (440-660 nm) [25]. The results were interpreted using a five-level model, which clearly demonstrated that the shorter wavelength region (440-560 nm) is dominated by the excited state absorption (ESA), whereas the longer wavelength region (580-660 nm) is mainly dominated by the two-photon absorption (TPA). This reports the advantage of using $C_{60}$ over the Porphyrin and Phthalocyanine due to the RAS behavior over the entire visible region. Kiran *et al* [26] reported the surface Plasmon-enhanced optical limiting behavior in Ag-Cu nanocluster co-doped in $SiO_2$ sol-gel film. The nonlinear optical properties of CdS were investigated for the potential usages as an optical limiter. The third-order nonlinear optical properties of different porphyrin samples were thoroughly investigated using femtosecond, picosecond, and nanosecond Z-scan technique. In 2010, Sathyavathi *et al* reported an eco-friendly and straightforward biosynthesis procedure of silver nanoparticles using Coriandrum sativum leaf extract [27]. The aqueous silver ions were reduced and result in silver in the presence of leaf extract. The average diameter of the synthesized silver nanoparticles was obtained around ~26 nm, which exhibited strong optical limiting behavior upon excitation of 532 nm laser pulse [27].

Supercontinuum generation is another exciting field nurtured by DNR. The interactions of intense ultrashort laser pulses with the transparent bulk medium leads to considerable modifications in the Spatio-temporal properties, which results in the generations of a white-light continuum ranging from ultraviolet to infrared. The large frequency sweep due to the propagation of ultrashort laser pulses is termed as supercontinuum generations caused by the different physical processes such as self-focusing, self-steepening, group velocity dispersion, anti-Stokes spectral broadening, intensity clamping, parametric four-photon mixing, competition between multiple filaments, etc. [28,29]. In 2005, Srinivas *et al* reported the first-ever femtosecond supercontinuum generation in a potassium di-hydrogen phosphate (KDP) using ~100 fs pulses at 790 nm irradiation. KDP shows an enhanced blue continuum depending on the angle of incidence [29]. Kumar *et al* [28] reported enhanced broadband supercontinuum generation in a potassium di-hydrogen phosphate (KDP). The enhancement and angle tunability of white light generation is demonstrated in the shorter wavelength (< 400 nm) regime employing the supercontinuum generation and second harmonic generation in tandem.

## 6 1D, 2D and 3D photonic crystals for photon confinement and optical field modulation

Another notable pioneering contribution of DNR and his group is in controlling of photons through artificial photonic crystals. From the past decade, many exciting research works of DNR can be found on one-(1D) and three-dimensional(3D) photonic crystals. In the following section, we give a brief review on the photonic crystals and the various linear and nonlinear optical studies done by DNR and his group.

In Photonic crystals light propagation show abnormal characteristics due to spatial periodic modulation of dielectric contrast in the order of light wavelength. The optical field confinement and the control of photons in the artificial photonic crystal structure can be achieved in a controlled manner. Photonic crystals can be fabricated in one-, two- and three-dimensional structures [Fig 6(a)] depending on the spatial periodicity of the constituents [30]. Phonon amplification was observed due to photon-plasmon interaction and enhanced optical field strength in a novel hybrid metal-dielectric microcavity structure [31]. Several orders of enhancement in Raman Stokes line intensity were observed due to photon-plasmon-phonon triplet interactions. The higher density of state at the photonic band edge leads to the enormous local field enhancement, which can be envisaged in enhancing weak Raman signal. The reduced group velocity at the photonic band edge results in a higher number of photonic density of states [32]. Chiasera *et al* reported an efficient protocol for the fabrication of $Er^{3+}$ activated 1D photonic crystal (distributed Bragg reflector, DBR) using the RF sputtering technique. The photoluminescence features were evidenced for the cavity effect on the emission behavior [33,34]. Enhanced nonlinear absorption and optical limiting behavior were reported in ZnO based 1D photonic crystal [35]. Strong local field confinement around the defect layer leads to the four times enhanced two-photon absorption coefficients with respect to the single layer of ZnO. Alee *et al* reported the synthesis of polystyrene (PS) nanospheres through emulsion polymerization technique for the fabrication of 3D photonic crystals. Vertical deposition method produces a good crystalline quality [36].

An enhanced nonlinear absorption behavior was reported by Alee *et al* in 3D polystyrene photonic crystal [37]. Alee *et al* reported the spectral and morphological changes of 3D polystyrene photonic crystals in the presence of alcohols [38]. A simple two-step fabrication was implemented to achieve inverse opal structure with prominent photonic bandgap properties [39]. Figures 6(b-d) show the linear optical properties of $BaTiO_3$ based 1D photonic crystal (optical microcavity) [40], where the photonic cavity mode is observed at 532 nm at an angle of 32º. The optical field confinements are prominent in the transverse electric field intensity ($|E|^2$) map [Fig 6(c)].

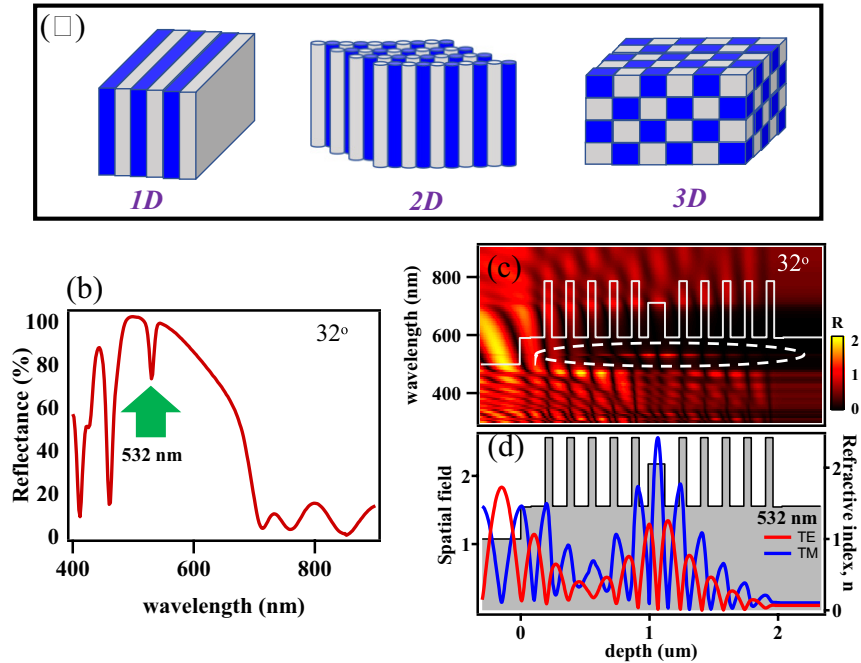

Fig 6. (a) Schematic representations of 1D, 2D, and 3D photonic crystals [30]. (b) Reflection spectrum of optical 1D microcavity realized from $BaTiO_3$ as defect layer sandwiched between two Bragg mirrors (of $SiO_2/TiO_2$). (c) Transfer matrix method (TMM) simulations for the transverse electric field ($|E|^2$) intensity map along with the spatial depth of the optical microcavity. (d) TMM simulations of spatial depth profile of transverse electric ($|E|$) and magnetic ($|H|$) fields.

Angle dependent nonlinear absorption enhancement and nonlinear absorption switching behavior were reported in $BaTiO_3$ based optical microcavity by Shihab *et al* [40,41]. Acharyya *et al* [23] reported the giant optical nonlinearity observed in $(Ag/SiO_2)_4$ metal-dielectric photonic structure. The silver intrinsic nonlinearity was enhanced manifold due to coupled *Fabry–Pérot* resonators.

In conclusion, a brief review on the pioneering contributions of Prof D Narayana Rao is highlighted in the context of optics and photonics research in India. Being an experimental physicist, the notion of optics to him is to realise the physical phenomenon with understandable and straightforward experiments. Due to fear of length, many of his notable contributions are intentionally not discussed here. Numerous researchers associated with him directly or indirectly are now well-placed in academia, both nationally and internationally. names of some of those researchers are given in Ref [42]

**Acknowledgements**

by him from the past several decades. We thank Dr B Maruti Manoj (IIT Kharagpur) for some inputs and Prof Soma Venugopal Rao (UoH) for editing the review.

films of poly-*p*-phenylenebenzobisthiazole polymer investigated by picosecond and subpicosecond degenerate four wave mixing, *Appl Phys Lett*, 48(1986)1187–1189.

22. Sheik-Bahae M, Said A A, Wei T H, Hagan D J, Van Stryland E W, Sensitive measurement of optical nonlinearities using a single beam, *IEEE J Quantum Elect*, 26(1990)760–769.

23. Acharyya J N, Rao D N, Adnan M, Raghavendar C, Gangineni R B, Prakash G V, Giant Optical Nonlinearities of Photonic Minibands in Metal–Dielectric Multilayers, *Adv Mater Interfaces*, 7(2020)2000035; doi.org/10.1002/admi.202000035.

24. Rao D N, Blanco E, Rao S V, Aranda F J, Rao D V G L N, Tripathy S, Akkara J A, A Comparative Study of $C_{60}$, Pthalocyanine, and Porphyrin for Optical Limiting Over the Visible Region, *J Sci Ind Res*, 57(1998)664–667.

25. Rao S V, Rao D N, Akkara J A, DeCristofano B S, Rao D V G L N, Dispersion studies of non-linear absorption in $C_{60}$ using Z-scan, *Chem Phys Lett*, 297(1998)491–498.

26. Kiran P P, Shivakiran Bhaktha B N, Rao D N, De G, Nonlinear optical properties and surface-plasmon enhanced optical limiting in Ag–Cu nanoclusters co-doped in $SiO_2$ Sol-Gel films, *J Appl Phys*, 96(2004)6717–6723.

27. Sathyavathi R, Krishna M B, Rao S V, Saritha R, Rao D N, Biosynthesis of silver nanoparticles using Coriandrum sativum leaf extract and their application in nonlinear optics, *Adv Sci Lett*, 3(2010)138–143.

28. Kumar R S S, Harsha S S, Rao D N, Broadband supercontinuum generation in a single potassium di-hydrogen phosphate (KDP) crystal achieved in tandem with sum frequency generation, *Appl Phys B*, 86(2007)615–621.

29. Srinivas N N, Harsha S S, Rao D N, Femtosecond supercontinuum generation in a quadratic nonlinear medium (KDP), *Opt Express*, 13(2005)3224–3229.

30. Joannopoulos J D, Johnson S G, Winn J N, Meade R D, Photonic Crystals: Molding the Flow of Light - 2nd Edn, (Princeton University Press), 2008.

31. Guddala S, Dwivedi V K, Prakash G V, Rao D N, Raman scattering enhancement in photon-plasmon resonance mediated metal-dielectric microcavity, *J Appl Phys*, 114(2013)224309; doi.org/10.1063/1.4842995.

32. Guddala S, Kamanoor S A, Chiappini A, Ferrari M, Rao D N, Experimental investigation of photonic band gap influence on enhancement of Raman-scattering in metal-dielectric colloidal crystals, *J Appl Phys*, 112(2012)084303; doi.org/10.1063/1.4758315.

33. Chiasera A, Jasieniak J, Normani S, Valligatla S, Lukowiak A, Taccheo S, Rao D N, Righini G C, Marciniak M, Martucci A, Ferrari M, Hybrid 1-D dielectric microcavity: fabrication and spectroscopic assessment of glass-based sub-wavelength structures, *Ceram Int*, 41(2015)7429–7433.

34. Chiasera A, Jasieniak J, Normani S, Valligatla S, Lukowiak A, Taccheo S, Rao D N, Righini G C, Marciniak M, Martucci A, Ferrari M, Fabrication and Spectroscopic Assessment of Glass-Based Sub-Wavelength Structures for Hybrid 1-D Dielectric 633-nm Laser Microcavity, in Advanced Solid State Lasers, OSA Technical Digest (online) (Opt Soc of Am), 2014; paper ATh2A.4.

35. Valligatla S, Chiasera A, Varas S, Das P, Bhaktha B S, Łukowiak A, Scotognella F, Rao D N, Ramponi R, Righini G C, Ferrari M, Optical field enhanced nonlinear absorption and optical limiting properties of 1-D dielectric photonic crystal with ZnO defect, *Opt Mater*, 50(2015)229–233.

36. Alee K S, Brundavanam M M, Bhaktha S N B, Chiappini A, Ferrari M, Rao D N, Effect of dye on the band gap of 3D polystyrene photonic crystals, Proc SPIE 7212, Optical Components and Materials VI, 72120R (6 February 2009); doi.org/10.1117/12.811001.

37. Alee K S, Krishna M B M, Ashok B, Rao D N, Experimental verification of enhanced electromagnetic field intensities at the photonic stop band edge of 3D polystyrene photonic crystals using Z-Scan technique, *Photonic Nanostruct*, 10(2012)236–242.

38. Alee K S, Sriram G, Rao D N, Spectral and morphological changes of 3D polystyrene photonic crystals with the incorporation of alcohols, *Opt Mater*, 34(2012)1077–1081.

39. Guddala S, Alee K S, Rao D N, Fabrication of multifunctional $SnO_2$ and $SiO_2$-$SnO_2$ inverse opal structures with prominent photonic band gap properties, *Opt Mater Express*, 3(2013)407–417.

40. Shihab N K, Acharyya J N, Rasi U M, Gangineni R B, Prakash G V, Rao D N, Cavity enhancement in nonlinear

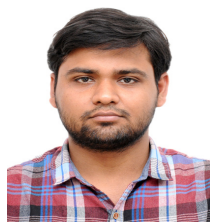

Jitendra Nath Acharyya received his M Sc degree in Physics in 2017, from Vidyasagar University, India. Currently he is a doctoral research student working under the supervision of Prof  G Vijaya Prakash, Department of Physics, Indian Institute of Technology Delhi, India from 2018 as DST-INSPIRE fellow. His research focuses on Nonlinear Optical Properties and Ultrafast Dynamics in Photonic Structures. The study comprises of the third-and higher-order nonlinear optical properties of microcavities and DBRs utilizing nonlinear optical techniques such as Z-scan and the ultrafast transient absorption spectroscopy utilizing the ultrafast high-power femtosecond lasers. He has  published a number of papers in journals of international repute.